\documentclass[12pt]{article}

\usepackage{epsfig}

\def\W{{\phi}}

\def\laH{{\lambda_{\rm H}}}
\def\laL{{\lambda_{\rm L}}}

\def\deH{{\rho_{\rm H}}}
\def\deL{{\rho_{\rm L}}}

\title{\vglue -0.5in
\bf {Two-phase densification of cohesive\\granular aggregates}}
\author{G.\ Gioia$^1$ and  A.\ M.\ Cuiti\~no$^2$ }  
\date{}
\begin{document}
\maketitle
\thispagestyle{empty}
\vskip -.8cm
{\noindent \scriptsize ${}^1$ 
Department of Theoretical \& Applied Mechanics,
University of Illinois,
Urbana, IL 61801, USA.}\\ 
{\noindent \scriptsize ${}^2$ 
Department of Mechanical \& Aerospace Engineering,           
Rutgers University,
Piscataway, NJ 08854, USA.}\\
\vskip 1mm
\begin{abstract}
{\bf \noindent 
When poured into a container, cohesive 
 granular materials form low-density, open granular 
 aggregates. If pressed upon with a ram, these aggregates 
 densify by particle rearrangement. 
 Here we introduce 
 experimental evidence to the effect
 that particle rearrangement is a spatially 
 heterogeneous phenomenon, which occurs in the form 
 of a phase transformation between two 
 configurational phases of the granular aggregate.
 We then show that the energy landscape associated with 
 particle rearrangement is consistent with 
 our interpretation of the experimental results.
  Besides affording insight into the physics of the granular 
 state, our conclusions are relevant to many 
   engineering processes and natural phenomena.}
\end{abstract}

\begin{figure}
\vskip -0in
\epsfxsize=5.4in
\centerline{\epsfbox{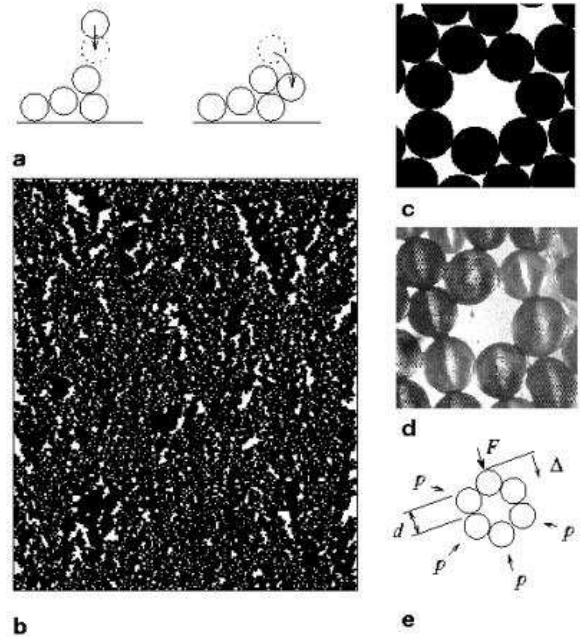}}
\caption{\small
 {\bf a)} Computer simulation of container filling 
  by a ballistic  aggregation method.$^8$ 
 The 
 particles are sequentially deposited along random
  vertical paths,   and then allowed to roll until they make contact 
  with any two points (at least) on the surface of the aggregate.
   The magnitude of the cohesive forces is assumed strong enough 
 to stabilize a particle both at the time of aggregation 
    and afterwards, but it remains otherwise unspecified. 
 {\bf b)} Obtained particle aggregate. 
 {\bf c)} More or less regular rings of particles 
   are  a pervasive feature of
 the simulated aggregate, and also of {\bf d)}
  aggregates obtained in quasi two-dimensional
   container-filling experiments. Note the water 
 menisci among the particles. 
   {\bf e)} Collapse of a ring of particles by 
 snap-through buckling: one of the particles jumps 
 to the center of the ring.  The driving force $F$ 
 is a relatively large
 contact force localized on the jumping
 particle;  both photo-elastic and  numerical 
 studies indicate$^5$  that such forces exist with magnitudes 
 several times larger than the average contact force, which 
 is represented by the hydrostatic pressure $p$ 
  }
\label{fig0}  
\end{figure}

\noindent Cohesive granular materials have been the focus of 
 a minute fraction of recent research into the granular 
 state.$^1$ 
 Yet cohesive granular materials will surely draw increasing 
 attention from scientists and engineers, 
 if only because they are used in numerous 
  applications.    Conspicuous  examples 
 are afforded by the forming of ceramic parts, pulvimetallurgic 
 components and pharmaceutical tablets by  compaction 
  of fine powders.$^2$ 
 The cohesiveness of powders
 stems from the large surface/volume 
  ratio of their constitutive particles, which 
  enhances the effect of attractive
  van der Walls forces. In other applications, 
 e.g.\ the stabilization of soils, the 
 cohesiveness is due to the presence 
  of liquid menisci among the particles. 
     Our interest in the densification 
 of cohesive granular materials was prompted by the 
  recent compaction 
 study of Kong and Lannutti.$^3$ 
 These authors used X-ray tomography to document the 
 evolution of density during the compaction of 
  alumina powders (diameter $\sim 60\,\mu$m). They reached the 
  tantalizing conclusion that densification 
 ``seems to proceed as a wave initiated at the  
 advancing ram''.$^3$ 
 Our aim here is to elucidate the nature of 
 this `wave,' and to relate its behavior to 
 the micromechanics of densification 
 in cohesive  granular materials.

\begin{figure}
\vskip .5in
\epsfxsize=5.48in
\centerline{\epsfbox{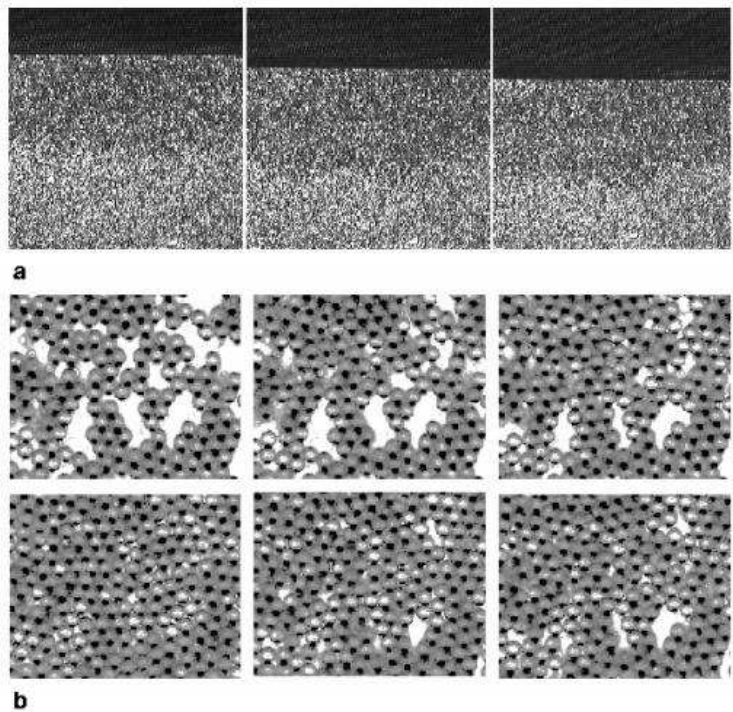}}
\caption{\small {\bf  a)} Three sucessive times during 
  densification.  The ram is black, and moves 
 from the top down. The high- and low-density regions 
 are dark and light gray, respectively. The area shown 
 is $\sim 20\times 20\,$cm$^2$.
   {\bf  b)} Clockwise starting from the top left:
 successive collapse of rings of particles 
 in a vicinity of the rearrangement front.  
  The particle diameter is 1.7$\,$mm.
  }
\label{expfot} 
\end{figure}

  When, preceding compaction, a cohesive
 granular material is poured into a container,
 the mobility of the particles reaching the
 bottom of the container is  hindered by the 
 cohesive forces, Fig.~\ref{fig0}a. As a result, a  
 low-density, open aggregate of particles  
 obtains inside the container, Fig.~\ref{fig0}b. 
 Open aggregates densify by  particle rearrangement
 at relatively low pressure.$^4$ 
  It has been proposed$^5$ 
 that particle rearrangement occurs when the
  {\it rings of particles\/} of the open aggregate 
 collapse by snap-through buckling, see 
 Fig.~\ref{fig0}c-e. To investigate this phenomenon 
  we prepared a quasi two-dimensional open aggregate$^6$ 
 by  filling a narrow plexiglass container 
  (thickness $\sim 1.9\,$mm)
  with monosized glass beads  (diameter $\sim 1.7\,$mm).  
 Before pouring them into the container, we wetted the beads with
  water in order for  menisci to form among 
  the beads. These menisci provided the required cohesion. 
 We then compacted the aggregate using 
 a ram. Fig.~\ref{expfot}a shows 
 three stages during the experiment.  
  A high-density region  (the phase H, wherein rearrangement
  has taken place already)  and a low-density region 
 (the phase L, wherein the open aggregate  

\begin{figure}
\vskip -.0in
\epsfxsize=5.48in
\centerline{\epsfbox{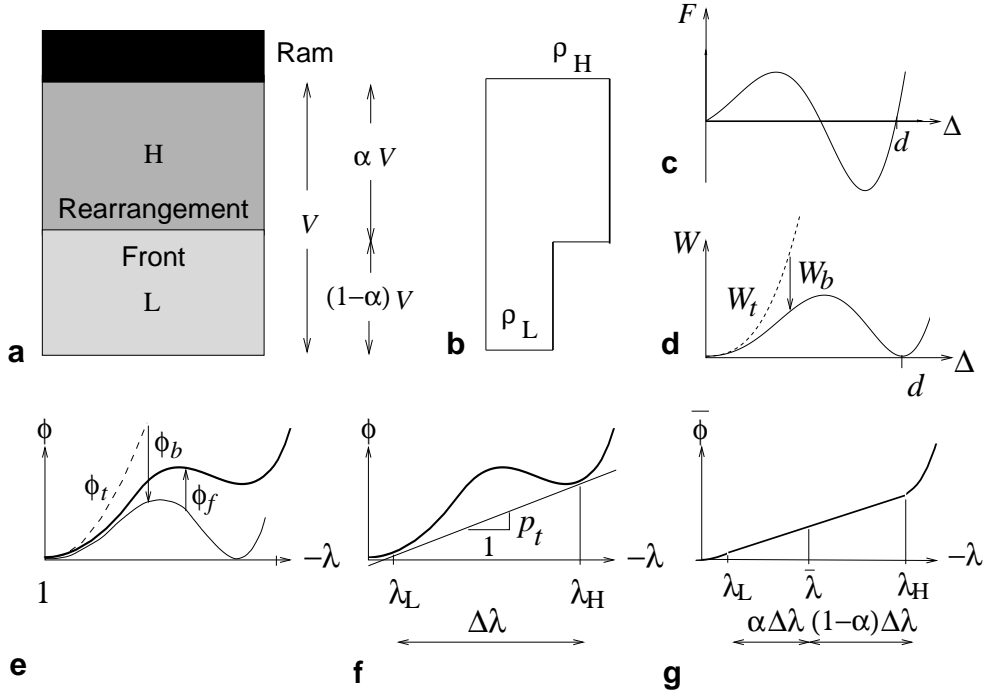}}
\caption{\small    {\bf  a)} Interpretation of the experimental 
 results:  {\it V} is the total volume of the aggregate; 
 the phases H and L are  separated by the rearrangement front; 
  $\alpha$ is the volume fraction of H.
  As $\alpha$  increases from 0 to 1, the rearrangement front 
 sweeps through the aggregate from ram to bottom. 
  Cf.\ Fig.~\ref{expfot}a. 
  {\bf b)} The  density jumps by  $ \Delta\rho = 
 \rho_{\rm H}-\rho_{\rm L}$ across the rearragement front.
  {\bf c) } Mechanical response of the ring of particles 
 of Fig.~\ref{fig0}e:
 force $F$ vs.\ displacement $\Delta$ (schematic). 
  $d$ is the particle diameter. 
 {\bf d)} Internal energy $W$ vs.\ $\Delta$.  
 {\bf e)}   We  express the energy per unit volume
 of aggregate  as $\W=\W_t+\W_b+\W_f$, where 
  $\W_t$ and $\W_b$ correspond to  
 the terms  $W_t$ and $W_b$ of (d)   
   averaged over a statistically representative 
  volume of aggregate, and  $\W_f$ is
  the energy per unit volume dissipated 
  by the inter-particle frictional forces.
  {\bf f)} Geometrical interpretation of the 
 equilibrium equations, Eq.~\ref{erdmann}. 
 {\bf g)} The average  energy per unit volume 
  ${\bar \W}({\bar \lambda})$.
}
\label{theofot} 
\end{figure}

 remains  unchanged) are clearly discernable.
  Densification proceeds
  by growth of the volume fraction of H  at the expense 
 of the volume fraction of L. Visual inspection revealed 
 that no  rearrangement occurs within the high- or 
 low-density regions. In a narrow vicinity
 of the H-L interface or  {\it rearrangement front\/},  
  on the other hand, we 
 could clearly observe the collapse by 
 snap-through buckling of successive layers of  rings
 of particles, see  Fig.~\ref{expfot}b. The rearrangement front,
  which we identify with 
 the densification wave of Kong and Lannutti, 
   advances in the form of broad, shallow
  tongues (visible in 
 Fig.~\ref{expfot}a) darting forward 
  in coincidence with the collapse of ring of particles.
  These results suggest that densification occurs 
 in the form of a phase transformation L$\rightarrow$H, 
  Fig.~\ref{theofot}a-b. To substantiate out interpretation
  of the experimental evidence we turn  our attention to the 
   micromechanics of particle rearrangement.
   Consider  a ring of particles undergoing snap-through 
 buckling, Fig.~\ref{fig0}e. For an increasing  
 displacement $\Delta$, Fig.~\ref{theofot}c,  
  the force $F$  vanishes as the particle snaps 
 into the void, and again when the particle reaches 
 the center of the void.  Fig.~\ref{theofot}d shows
  the attendant evolution of internal energy. 
 The energy associated with the initial (tangent) response 
 of the ring, $W_t$, is represented by a dashed curve; 
 the relaxation effected by buckling, $W_b$, is 
 represented by an arrow. 
 $W_b$ causes the function  $W(\Delta)$  to be   nonconvex. 
 Nonconvex energy functions are  characteristic of 
 systems which undergo phase transformations.$^7$  

 To move on  to the  macroscopic scale we consider 
 an open granular aggregate contained in a 
 {\it frictionless\/} container of volume $V$ and 
 constant cross-sectional area. 
  We study the energetics of densification 
 in the space of the local stretch $\lambda$.
 (The local stretch is defined by $\lambda=\rho_0/\rho$, 
 where $\rho$ is the local density and $\rho_0$ is
  the density of the initial open aggregate.)
  The energy per unit volume 
   is $\W(\lambda)$, Fig.~\ref{theofot}e; it is nonconvex by
 inheritance from $W(\Delta)$, Fig.~\ref{theofot}d.
  The local pressure 
 is given by $p=-d\W(\lambda)/d\lambda$. We now set the average 
 stretch to a given value, 
  ${\bar\lambda} ={{1}\over{V}}\int_V \lambda dV$,   and  
 minimize the total energy of the 
  aggregate, $\int_V \W(\lambda) dV$,  
 using conventional tools of nonconvex analysis.$^7$ 
 This leads to the following equilibrium equations
\begin{equation}
\label{erdmann}
p_t= - {{\W(\laL)-\W(\laH)}\over{\laL-\laH}} = -{{d\W}\over{d\lambda}}(\laL) = -{{d\W}\over{d\lambda}}(\laH),
\end{equation}
 which allow for the computation of the {\it characteristic stretches\/} 
 $\laL$ and $\laH$ and the {\it transformation pressure\/}  $p_t$,
  Fig.~\ref{theofot}f. It is apparent from Eq.~\ref{erdmann} that 
 $\laL$, $\laH$ and $p_t$ are independent of $\bar\lambda$, and can 
 be construed as material properties. The characteristic stretches 
 define two configurational 
 phases L and H of density $\deL=\rho_0/\laL$ and 
 $\deH=\rho_0/\laH$, respectively.  When $\laL > {\bar \lambda} > \laH$, 
 the phase H occupies a volume $\alpha V$, and the phase L 
 a volume $(1-\alpha) V$, where 
 $\alpha=(\laL-{\bar \lambda})/(\laL-\laH)$,
  Fig.~\ref{theofot}a. As $\alpha$ increases from 
 0 to 1 during densification, 
 ${\bar \lambda}$ decreases fom $\laL$ to $\laH$, the average
 density  increases from $\deL $ to $\deH $, and the 
 pressure remains equal to $p_t$.   It is noteworthy 
 that $p_t \neq 0$ on account of the dissipative effects 
 of inter-particle friction, Fig.~\ref{theofot}e-f.   
 The average energy per unit volume is given by the  convexified form 
${\bar \W}= \alpha \W(\laH)+(1-\alpha)\W(\laL)$,
 which for $\laL > {\bar \lambda} > \laH$ 
 yields the correct value $p=-d{\bar\W}(\bar\lambda)/d{\bar\lambda}=p_t$.

   The spatial distribution of phases  is 
 not given by our analysis. Nuclei of H could form  
 in many places throughout the aggregate,  
 leading to  stratified  mixtures of H and L. There is 
 no indication of such mixtures in Fig.~\ref{expfot}a,
 however.  We ascribe this fact 
 to the regularizing  effect of container-wall roughness.
  Because of wall roughness, the stress 
 decreases monotonically away from the ram. 
 Nucleation must  therefore occur at the ram, leading to
 the establishment of  a single H-L interface---the 
 rearrangement front. We will explore other effects 
 of wall roughness in a separate publication. 

 
 We conclude that if, as indicated by 
 our  experimental observations,
   snap-through buckling is 
 the dominant mechanism of  particle rearrangement, 
 then densification must occur in the form of a 
 phase transformation L$\rightarrow$H. 
 Container-wall roughness  accounts for the presence of a 
 single  H-L interface. 
  It follows that our interpretation of
  the experimental evidence is consistent
 with the micromechanics of particle rearrangement.  

 \newpage

\subsection*{\bf References }

\setlength{\itemsep}{0.ex}
\frenchspacing        
\begin{enumerate}

\item Jaeger, H.\ M.\ and  Nagel, S.\ R.
 Physics of the granular state.
 {\it Science} {\bf 255}, 1523--1531 (1992).

\item Ewsuck, K.\ G.
 Compaction science and technology. 
 {\it MRS Bull.\/} {\bf 22}, 14--16 (n.\ 12 1997).

%

\item Kong, C.\ M.\ and  Lannutti, J.\ J.
 Localized densification during the compaction of alumina granules: 
 the stage I--II transition.
  {\it J.\ Am.\ Ceram.\ Soc.\/} {\bf 83}, 685--690 (2000).

\item Niesz, D.\ E., Bennett, R.\ B.\ and Snyder, M.\ J.
 Strength characterization of powder aggregates.
 {\it Ceram.\ Bull.\/} {\bf 51}, 677-680 (no.\ 9 1972). 

\item Kuhn, L.\ T.\,  McMeeking, R.\ M. and Lange, F.\ F.
 A model for powder consolidation.
{\it J.\ Am.\ Ceram.\ Soc.\/} {\bf 74}, 682--685 (1991).

\item Jaeger, H.\ M., Knight, J.\ B., Liu, C.,  and  Nagel, S.\ R.
 What is shaking in the sandbox?
 {\it MRS Bull.\/} {\bf 19}, 25--31 (n.\ 5 1994).


%

\item Ericksen, J.\ L., {\it Introduction to the thermodynamics  
 of solids\/}, 2nd edition, ch.\ 3. (Springer-Verlag, New York, 1998).

\item Jullien, R.\ and  Meakin, P.
 Simple three-dimensional 
 models for ballistic deposition with restructuring.
{\it Europhys. Lett.\/} {\bf 4}, 1385--1390 (1987).

\end{enumerate}

\noindent  Akcknowledgements. 
 Supported by the International 
 Fine Particle Research Institute and 
the New Jersey Commission on Science and Technology. 
We thank Mr.\ Tom\'as Uribe for his help with the experimental work. 


\newpage

\end{document}